\documentclass[aps,pra,onecolumn]{revtex4}

\usepackage[usenames,dvipsnames]{color}
\usepackage[latin1]{inputenc}
\usepackage{graphicx}
\usepackage{amssymb}
\usepackage{amsmath}
\usepackage{amsfonts}
\usepackage{bm}
\usepackage{ulem}
\usepackage{color}
\newcommand{\ket}[1]{\ensuremath{\left|{#1}\right\rangle}}
\newcommand{\bra}[1]{\ensuremath{\left\langle{#1}\right |}}
\newcommand{\sd}{\ensuremath{\downarrow}}
\newcommand{\su}{\ensuremath{\uparrow}}

\newcommand{\beq}{\begin{equation}}
\newcommand{\eeq}{\end{equation}}
\newcommand{\bea}{\begin{eqnarray}}
\newcommand{\eea}{\end{eqnarray}}
\newcommand{\be}{\begin{equation}}
\newcommand{\ee}{\end{equation}}
\newcommand{\ben}{\begin{eqnarray}}
\newcommand{\een}{\end{eqnarray}}

\begin{document}

\title{Entanglement generation through particle detection in systems of identical fermions}

\author{P. A. Bouvrie$^{1}$, A. Vald\'es-Hern\'andez$^{2}$, A. P. Majtey$^{3,4}$, C. Zander$^5$, and A. R. Plastino$^{4,6}$}

\affiliation{$^1$Centro Brasileiro de Pesquisas F\'isicas, Rua Dr. Xavier Sigaud 150, Rio de Janeiro, RJ 22290-180, Brazil}
\affiliation{$^2$Instituto de F\'{\i}sica, Universidad Nacional Aut\'{o}noma de M\'{e}xico, \\
Apartado Postal 20-364, M\'{e}xico, Distrito Federal, Mexico}
\affiliation{$^3$Facultad de Matem\'atica, Astronom\'{\i}a y F\'{\i}sica, Universidad Nacional de C\'ordoba, Av. Medina Allende s/n, Ciudad Universitaria, X5000HUA C\'ordoba, Argentina}
\affiliation{$^4$Consejo de Investigaciones Cient\'{i}ficas y T\'ecnicas de la Rep\'ublica Argentina, Av. Rivadavia 1917, C1033AAJ, Ciudad Aut\'onoma de Buenos Aires, Argentina}
\affiliation{$^5$Physics Department, University of Pretoria, Pretoria 0002, South Africa}
\affiliation{$^6$CeBio  y Secretar\'{i}a de Investigaciones, Universidad Nacional del Noroeste de la Prov. de Buenos Aires, UNNOBA-Conicet, Roque Saenz-Pe\~{n}a 456, Junin, Argentina
}

\email[]{bouvrie@ugr.es, andreavh@fisica.unam.mx, amajtey@famaf.unc.edu.ar, cz@up.ac.za, arplastino@unnoba.edu.ar}

\begin{abstract}
We investigate the generation of entanglement in systems of identical fermions through
a process involving particle detection, focusing on the implications that this kind of processes have 
for the concept of entanglement between fermionic particles.  As a paradigmatic example  we discuss in detail a scheme 
based on a splitting-plus-detection operation. This scheme generates states with accessible 
entanglement starting from an initial pure state of two indistinguishable fermions exhibiting correlations due purely to 
antisymmetrization. It is argued that the proposed extraction of entanglement does not contravene the notion 
that entanglement in identical-fermion systems requires correlations beyond those purely due to their indistinguishability.
In point of fact, it is shown that this concept of entanglement, here referred to as {\it fermonic entanglement},
actually helps to clarify some essential aspects of the entanglement generation process. In particular, we prove that the amount
 of extracted accessible entanglement equals the amount of fermionic entanglement  created with the detection process. 
 The aforementioned scheme is generalized for the case of $N$-identical fermion systems of 
 arbitrary dimension. It transpires from our present discussion that a proper analysis of entanglement 
 generation during the splitting-plus-detection operation is not only consistent with
 the concept of fermonic entanglement, but actually reinforces this concept. 
 
 \vskip 0.3cm
 

\end{abstract}


\maketitle

\section{Introduction}
The most distinctive feature of quantum systems composed of $N$ identical fermions is that their elementary (antisymmetric) 
pure state is a single Slater determinant, hence it formally looks entangled (in contrast, for example, with a separable pure state of distinguishable qubits). 
The ensuing correlations, which we will refer to as `Slater correlations', can manifest themselves in one or more degrees of freedom. 
In the present contribution we investigate some aspects of entanglement generation schemes in fermion systems, 
emphasizing the light they shed on the nature of entanglement between identical fermions and, in particular, on the status  of the above mentioned minimal fermion correlations. 
There is widespread consensus that  entanglement between identical fermions is associated with the quantum correlations exhibited \textit{on top} of the Slater correlations
\cite{GMW02,ESBL02,PZ02,YS03,GM04,LNP05,GM2005,NV07,PMD09,BPCP08,ZP10,AFOV08,PY01,GM05,DDW06,LZLL01,TMB11}. 
This conception of entanglement in fermion systems is relevant, for instance, in atomic physics \cite{KO2014,ELD2015}, 
in quantum chemistry \cite{MELD2015}, and in the study of quantum dots \cite{SN2015}. From this perspective, 
a single Slater determinant should be considered as a non-entangled state, thus leading to the notion of \textit{fermionic entanglement} 
to account for the extra correlations beyond those due to the indistinguishability of the parties, and to the antisymmetric property of fermionic states. 
However, various proposals have been made where accessible entanglement is ``extracted" from minimally (Slater) correlated fermion systems  
(by accessible entanglement we mean entanglement shared between spatially separated entities, e.g. Alice and Bob). 
Indeed, there is an interesting ongoing debate over whether it is possible to use the strictly spin-statistical correlations 
associated with Slater states to perform quantum information tasks \cite{LNP05,PMD09,Plenio,CMMTF07}.
Here we analyze a scheme to generate accessible entanglement from the Slater correlations by means of a splitting-plus-detection operation. 
Specifically, we consider a system of two electrons located in a double-well potential. An initial single Slater determinant is subjected to a 
tunneling (or splitting) operation followed by a projective measurement.  The spin degrees of freedom of the projected indistinguishable-fermion 
state are then effectively described by an accessible entangled state of two distinguishable-qubits. We show that such entanglement can 
be generated only because fermionic entanglement is also created in the detection process. The measurement process and the interaction 
of the fermionic system with the measuring apparatus is analyzed with the aid of a simple though clarifying model, and a generalization of the splitting process to include systems of $N$ fermions is developed.\\

Even though most of our present considerations are based on a particular process, the concomitant analysis 
contributes to elucidate general features of fermionic entanglement. In particular, it highlights the fact that
the amount of accessible entanglement contained in a pure  two-fermions state (with the two particles localized
at spatially separated locations) is equal to the amount of fermonic entanglement.
In contrast with previous discussions on entanglement generation with identical particles  \cite{Plenio}, our analysis clearly identifies the detection process as the origin of the generated fermionic entanglement, and goes in line with the statement that the Slater correlations do not provide on their own a resource for implementing quantum information processing tasks \cite{GMW02,ESBL02,GM04,NV07}. This is consistent with the possibility of assigning complete sets of properties to the subsystems of the composite system \cite{GMW02,GM04}.\\

This work is structured as follows. In Sec. II we present a brief review of the definition of entanglement in $N$ identical fermion systems. In Sec. III we propose and analyze a scheme of entanglement extraction for systems composed by two identical fermions. We do this by considering a systems of two electrons located in a double-well potential and performing a splitting-plus-detection process acting on an initial non-entangled (in the fermionic sense) state. In Sec. IV we introduced an idealized toy model in order to analize the role of the detection operation. The process for generating useful entanglement is generalized to systems of $N$ identical fermions in a double-well potential in Sec. V. Finally, some conclusions are drawn in Sec. VI.

\section{Entanglement in systems of identical fermions}
Let us consider a system of $N$ identical fermions, and denote with $\{\ket{1},\ket{2},\ldots, \ket{d}\}$ a basis of the single-particle Hilbert space $\mathcal{H}_f$, of dimension $d\geq N$. The antisymmetric combination in $\mathcal{H}_f^{\otimes N}$
\bea
\label{slater}
\ket{\psi^{\textit{sl}}} = \frac{1}{\sqrt{N!}}\!\!\!\! \sum_{\substack{\{i_1,\ldots,i_N\}\\ \in S_p(1,\ldots,N)}} \!\!\!\!\!\varepsilon^{i_1\ldots i_N} \ket{i_{1}, i_{2}, \ldots, i_{N}},
\eea
defines what is called a Slater determinant. Here $\varepsilon^{i_1\ldots i_N}$ stands for the $N$-dimensional Levi-Civita tensor, and $S_p(1,\ldots,N)$ are the $N!$ permutations of the set $\{1,2,\ldots,N\}$, which correspond (without loss of generality) to the first $N$ elements of the $\mathcal{H}_f$ basis. An $N$-identical-fermion state is regarded as separable if and only if its density matrix is of the form \cite{GMW02}
\begin{equation}
\rho^{sep}=\sum_{k}p_{k}\ket{\psi^{sl}_k}\bra{\psi^{sl}_k}
\label{rhosep},
\end{equation}
with $\sum_{k}p_{k}=1.$ A state that cannot be decomposed as (\ref{rhosep}) is regarded as entangled in the fermionic sense (or endowed with fermionic entanglement).\\

\section{Entanglement extraction in two-fermion systems}
\label{EEtwofermions}
We start by considering a system of two electrons located in a double-well potential (e.g., a pair of electrons in coupled quantum dots \cite{SLM01}). The qubits are realized by the spin degree of freedom of the electrons, with states $\ket{\sd}$ and $\ket{\su}$. Let $\ket{A}$ and $\ket{B}$ denote the spatial part of the state, corresponding to spatially localized wave functions in the left ($A$) and right ($B$) well, with $\langle A|B\rangle=0$. Then, an orthonormal basis of the (four-dimensional) single-particle state space is $\{\ket{A}\ket{\sd},\ket{A}\ket{\su},\ket{B}\ket{\sd},\ket{B}\ket{\su}\}$. Initially the two electrons are in the left well, hence
\begin{equation}
\label{initState}
|\psi_{\text{init}}\rangle=\frac{1}{\sqrt{2}}(|A\rangle\ket{\sd}\otimes|A\rangle\ket{\su}
-|A\rangle\ket{\su}\otimes|A\rangle\ket{\sd}).
\end{equation}
Next, the potential barrier is reduced during a time interval $\tau$, leading to a non-vanishing tunneling amplitude $\sqrt{p}$, with $p=p(\tau) \in [0, 1]$ \cite{Strzys2010,Tichy2013}. Then, the potential barrier is raised again. The complete process is thus equivalent to a splitting transformation $U_{\textrm{split}}=U_f\otimes U_f$,
where the unitary operator $U_f$ acting on $\mathcal{H}_f$ is such that
\begin{eqnarray}
\label{unitary}
U_f|A\rangle\ket{\sigma}&=&\sqrt{1-p}|A\rangle\ket{\sigma}+\sqrt{p}|B\rangle\ket{\sigma},
\end{eqnarray}
where $\sigma$ is either $\su$ or $\sd$. The final state reads
\begin{eqnarray}
\label{finalstate}
 |\psi_{\text{final}}\rangle&=&\frac{1}{\sqrt{2}}[(1-p) (|A\rangle\ket{\sd}\otimes|A\rangle\ket{\su}-|A\rangle\ket{\su}\otimes|A\rangle\ket{\sd})+\sqrt{p(1-p)}(|A\rangle\ket{\sd}\otimes|B\rangle\ket{\su}-|B\rangle\ket{\su}\otimes|A\rangle\ket{\sd}\nonumber\\&+&|B\rangle\ket{\sd}\otimes|A\rangle\ket{\su}-|A\rangle\ket{\su}\otimes|B\rangle\ket{\sd})+p(|B\rangle\ket{\sd}\otimes|B\rangle\ket{\su}-|B\rangle\ket{\su}\otimes|B\rangle\ket{\sd})].
\end{eqnarray}
It is straightforward to verify that $|\psi_{\text{final}}\rangle$ is a single Slater determinant,
\begin{equation}
  |\psi_{\text{final}}\rangle=\frac{1}{\sqrt{2}}(|\xi_\sd \rangle\otimes|\xi_\su\rangle-|\xi_\su\rangle\otimes|\xi_\sd\rangle),
\end{equation}
where $|\xi_\sigma\rangle=U_f|A\rangle\ket{\sigma}$.
This confirms that no entanglement between the particles was created by the splitting operation, and is consistent with the fact that the splitting transformation on the two-fermion system is a unitary operation that is local in $\mathcal{H}_f\otimes \mathcal{H}_f$ (yet clearly is a spatially-nonlocal operation.) \cite{ESBL02}.

The terms proportional to $(1-p)$ and $p$ in Eq. (\ref{finalstate}) correspond to a situation in which both fermions are in the same well, and have no fermionic entanglement. Yet the term proportional to $\sqrt{p(1-p)}$  exhibits a finite amount of fermionic entanglement. By projecting (\ref{finalstate}) onto the state with one particle in each well, we obtain (after normalization)
\bea
\label{projected}
 \ket{\psi_{\text{proj}}}&=&\frac{1}{2}(|A\rangle\ket{\sd}\otimes|B\rangle\ket{\su}-|B\rangle\ket{\su}\otimes|A\rangle\ket{\sd}+|B\rangle\ket{\sd}\otimes|A\rangle\ket{\su}-|A\rangle\ket{\su}\otimes|B\rangle\ket{\sd}).
\eea
This state cannot be written as a single Slater determinant in any basis, hence it is entangled in the fermionic sense. If now we let an agent (e.g. Alice) in $A$, and an agent (e.g. Bob) in $B$ to have access to the particle in their corresponding well, then Alice and Bob each have a single qubit. These qubits are clearly distinguishable, since each one pertains to a distinguishable physical agent. Therefore, once we ascribe the basis $\mathcal{H}_A=\{\ket{A}\ket{\sd},\ket{A}\ket{\su}\}$ to Alice, and the basis $\mathcal{H}_B=\{\ket{B}\ket{\sd},\ket{B}\ket{\su}\}$ to Bob, the two-indistinguishable-fermion state (\ref{projected}) becomes effectively equivalent to the two-distinguishable-qubit state,
\be
\label{projectedDisting}
\ket{\psi}=\frac{1}{\sqrt{2}}(\ket{\sd}_A\otimes \ket{\su}_B-\ket{\su}_A\otimes \ket{\sd}_B),
\ee
which is a maximally entangled Bell state \cite{ESBL02}. The indices denote Alice's and Bob's particles, and since both are localized and individually accessible ---unlike the  particles in (\ref{initState})---, $\ket{\psi}$ can be used as a resource for non-trivial quantum information tasks.

The useful (mode) entanglement in (\ref{projectedDisting}) reflects the fermionic entanglement contained in \eqref{projected} \cite{ESBL02}. In its turn,  \eqref{projected} arose as a result of a detection process, as will be described in more detail below. Thus it is possible to generate states endowed with accessible entanglement starting from a single Slater determinant, in a way that is consistent with the definition of fermionic entanglement explained above. In fact, we can resort to the (fermionic) concurrence introduced in \cite{ESBL02}, to quantify the entanglement between the fermions in Eq. (\ref{projected}). The result is that $\ket{\psi_{\textrm{proj}}}$ is a maximally entangled state, so $\ket{\psi_{\textrm{proj}}}$ and $\ket{\psi}$ have the same amount of entanglement, but    pertaining    to subsystems of different nature. It is important to note that even though the state obtained after the splitting (and before detection) has spatial entanglement, it still does not have accessible entanglement because it does not have a definite number of particles in each of the two spatial locations. In order to obtain a state with accessible entanglement (allowing for the implementation of standard protocols like teleportation), one needs also to perform the detection process which, besides leading to spatial accessible entanglement, also generates fermionic entanglement.

In the general case of two identical fermions
with a $(2n)$-dimensional single-particle Hilbert space, a particle-detection
process resulting in one particle located at $A$ and the other one located at $B$, leads (irrespective of the measured state)
to a state of the form
\begin{equation}
\label{locatab}
|\psi\rangle=\sum_{i,j=1}^n \frac{c_{i,j}}{\sqrt{2}}(|A\rangle |i\rangle \otimes |B\rangle  |j\rangle
-  |B\rangle |j\rangle \otimes |A\rangle  |i\rangle),
\end{equation}
where the labels $i,j$ correspond to the states of the internal degrees of freedom of the fermions,
 $\{|i\rangle, i=1, \ldots, n\}$ denotes an orthonormal basis for the associated
$n$-dimensional Hilbert space (for instance, for $s$-spin fermions we have $n=2s+1$),
and $c_{i,j}$ are complex coefficients with $\sum_{i,j} |c_{i,j}|^2 =1$.
When agents located at $A$ or $B$ operate only upon the internal degree of freedom,
the two-fermion state is effectively described by the state
 \begin{equation}
\label{locatabeff}
  |\psi\rangle_{\rm eff}=\sum_{i,j=1}^n c_{i,j}|i\rangle_A \otimes |j\rangle_B
\end{equation}
 of a bipartite system consisting of two distinguishable subsystems,
 each one with an $n$-dimensional Hilbert space.

 If we now consider the Schmidt decomposition
 of (\ref{locatabeff}),
 \bea
 |\psi\rangle_{\rm eff}=\sum_{i=1}^n
 \sqrt{\lambda_i}|\alpha_i\rangle_{A} \otimes |\beta_i\rangle_{B},
 \eea
 it is verified
 after some algebra that the fermionic Schmidt-like decomposition of the state (\ref{locatab})
 is given by
 \bea
 |\psi\rangle=\!\!\sum_{i=1}^n \!\sqrt{\frac{\lambda_i}{2}}
 (|A\rangle |\alpha_i\rangle \otimes |B\rangle|\beta_i\rangle \!-\!
 |B\rangle|\beta_i\rangle \otimes |A\rangle |\alpha_i\rangle).
 \eea
 It follows from
 these Schmidt decompositions that the quantitative amount of entanglement of the effective state
 (\ref{locatabeff}) coincides with the {\it fermionic} amount of entanglement exhibited by
 the full two-fermion state (\ref{locatab}). Using the linear entropy $S=1-\textrm{Tr}\rho^{2}_f$ of the
 single-particle reduced density matrix $\rho_f$ to quantify the amount of entanglement,
we get $S=1-\sum_i \lambda_i^2$ both for the full fermionic state (\ref{locatab}) 
and for the effective  state  (\ref{locatabeff}), a quantity that vanishes if the state
 (\ref{locatab}) is given by a single Slater determinant. Therefore, when
 a particle-detection process results in a pure state of two fermions localized at different places (with
 spatially non-overlapping wavefunctions), it is impossible to generate useful entanglement
 without at the same time generating fermionic entanglement. Indeed, the amount of useful entanglement
 obtained is precisely the same as the amount of fermionic entanglement generated.

Comparison of  (\ref{initState}) and (\ref{projectedDisting}) suggests that the net effect of the splitting-plus-detection process is that of transforming the Slater correlations between the individually unaccessible  fermions into entanglement correlations between two independently accessible qubits. We must stress, however, that this interpretation has to be taken with a grain of salt, because the original state (\ref{initState}) is non-entangled, whereas the two-fermion state (\ref{projected}) resulting from the detection operation is entangled. It is plain that the finite amount of entanglement exhibited by (\ref{projected}) was created during the detection process: it was not originally contained in the state (\ref{initState}).

It is worthwhile to mention that although for operational purposes one is ultimately interested in the state (\ref{projectedDisting}), it is possible to experimentally certify that the full two-fermion
state created after the detection process is actually (\ref{projected}). In order to do so, once
the projective measurement is performed on    the state (\ref{finalstate}) ---thus obtaining a state of the form (\ref{projected})---, we repeat the splitting transformation using Eq. (\ref{unitary}) and $U_f|B\rangle\ket{\sigma}=
\sqrt{1-p}|B\rangle\ket{\sigma}-\sqrt{p}|A\rangle\ket{\sigma}$  \cite{Strzys2010,Tichy2013}. Then,
the probability of finding both fermions in the same mode is found to be
$\mathcal P_{AA}=\mathcal P_{BB}=2p(1-p)$,
whereas the probability of finding a particle in each mode is $\mathcal P_{AB}=1-4p(1-p)$.
If, in contrast, the splitting transformation were performed on a state of the form (\ref{projectedDisting}) the resulting probabilities would be $\mathcal P_{AA}=\mathcal P_{BB}=p(1-p)$, and $\mathcal P_{AB}=1-2p(1-p)$.
Therefore, by experimental determination of the counting statistics of particles in both modes it can be certified that the state
created is (\ref{projected}).\\

\section{Interaction of the fermionic system with the measuring apparatus}
A fundamental feature of systems of identical parties is that the allowed transformations over the system preserve the exchange-symmetry of the state. An immediate consequence is that when a measurement is performed on a two-fermion system, the interaction between the fermions and the measurement apparatus
(or detector) affects each particle
in the same way. Thus the Hamiltonian $H_{\rm int}$
describing the interaction of the two fermions with the apparatus ${\cal M}$ has the form \cite{VHMP2015}
\begin{equation}
\label{hamintermea}
H_{\rm int} = H_{fM} \otimes \mathbb I_f + \mathbb I_f \otimes H_{fM}.
 \end{equation}

\noindent
Here $\mathbb I_{f}$ denotes the identity operator acting on the single-particle Hilbert space ${\cal H}_{f}$, and $H_{fM}$ describes the interaction between one fermion and the detector, and acts on the Hilbert space ${\cal H}_{f}\otimes {\cal H}_M$, where ${\cal H}_M$ is the Hilbert space associated with the detector.

In the problem at hand the aim of the measurement is to determine how many
fermions are, let us say, in the well $A$. We can consider an idealized toy model
that captures the essence of the situation. We start by analyzing the interaction between a single fermion and the measuring apparatus. Let us assume that ${\cal H}_M$ has an orthonormal basis
$\{|n\rangle\}$, with $n=0, \pm 1, \pm 2, \ldots$. When the fermion is
in the well $A$, and its interaction with the detector is turned on
during a time interval $\tau$, the resulting evolution, governed by the unitary operator $U_{\rm int} = \exp \left(-i H_{\rm int} \tau/\hbar \right)$, leaves the state of the particle unchanged and changes the state of the detector
according to the unitary transformation $|n \rangle \rightarrow |n + 1\rangle$.
The fermion-detector interaction is local: if the particle
is in the well $B$ it does not interact with the detector. Therefore, the interaction changes
the state of the detector according to the location of the particle, leaving
the state of the particle unaffected (this situation resembles the CNOT gate). If $|\psi_A \rangle=|\psi \rangle |A \rangle$ is a single-fermion state with the particle localized in the
well $A$, $H_{fM}$
acts according to
\beq
H_{fM} |\psi_A\rangle |n\rangle = |\psi_A\rangle \left( \sum_k c_{n,k} |k\rangle \right),
\eeq
where the coefficients $\{c_{n,k}\}$ form an appropriate Hermitian matrix. For a single-fermion state
$|\psi_B \rangle=|\psi \rangle |B \rangle$ corresponding to a fermion localized in the well $B$, we have $H_{fM} \ket{\psi_B} \ket{n} = \epsilon_0 \ket{\psi_B} \ket{n}$. We choose the zero of energy such that $\epsilon_0 = 0$. Then, we have
\bea
H_{fM}=\ket{A}\bra{A}\left(\sum_{k,n} c_{n,k}\ket{k}\bra{n}\right).
\eea

We now turn to the case in which the detector interacts with a two-fermion system, so that the interaction Hamiltonian is of the form \eqref{hamintermea}. The initial state of the detector is $\ket{0}$; the interaction is then turned on during a time interval $\tau$.
If both fermions are localized at the well $B$ (as, for example, in the state $|\Phi_0\rangle =\frac{1}{\sqrt{2}}(\ket{B}\ket{\downarrow}\otimes \ket{B}\ket{\uparrow}-  \ket{B}\ket{\uparrow}\otimes \ket{B}\ket{\downarrow}$), one has $U_{\rm int} \ket{\Phi_0} \ket{0} = \ket{\Phi_0} \ket{0}$. If there is only one fermion in the well $A$ (as happens, for example, in the state $|\Phi_1\rangle =\frac{1}{\sqrt{2}}(\ket{A}\ket{\downarrow}\otimes \ket{B}\ket{\uparrow}-  \ket{B}\ket{\uparrow}\otimes \ket{A}\ket{\downarrow}$), it can be verified that $H_{\rm int}$ satisfies $H_{\rm int} |\Phi_1\rangle |n\rangle = |\Phi_1\rangle \left( \sum_k c_{n,k} |k\rangle \right)$. Therefore, the associated time evolution operator $U_{\rm int}$ yields a transformation similar to the one obtained when having only one particle interacting with the detector, and $U_{\rm int}|\Phi_1\rangle |0\rangle=|\Phi_1\rangle |1\rangle$.
If instead the two fermions are in the well $A$ (as, for example, in the state $|\Phi_2\rangle =\frac{1}{\sqrt{2}}(\ket{A}\ket{\downarrow}\otimes \ket{A}\ket{\uparrow}-  \ket{A}\ket{\uparrow}\otimes \ket{A}\ket{\downarrow}$), we have $H_{\rm int} |\Phi_2\rangle |n\rangle = |\Phi_2\rangle \left( \sum_k 2 c_{n,k} |k\rangle \right)$. In this case, the time evolution operator corresponding to $H_{\rm int}$ and a time interval $\tau$  has the same effect upon the detector as the evolution operator corresponding to $H_{fM}$ and a time interval $2 \tau$. This effect is therefore the same as that of applying twice the transformation $|n\rangle \rightarrow |n+1\rangle$, yielding $|n\rangle \rightarrow |n+2\rangle$. In summary, the resulting transformation is $U_{\rm int}|\Phi_2\rangle |0\rangle=|\Phi_2\rangle |2\rangle$. In all cases the number of particles localized in the well $A$ can be read in the final state of the detector.

Consider now a two-fermion state that does not have a definite
number of particles in the well $A$. An example of such state is $|\psi_{\text{final}}\rangle$, given by Eq. (\ref{finalstate}). According to the above, the interaction between the fermions and the measurement apparatus leads to $U_{\rm int}|\psi_{\text{final}}\rangle |0\rangle=|\Psi\rangle$, with
\bea
\label{transf}
|\Psi\rangle&=&\frac{(1-p)}{\sqrt{2}}(\ket{A}\ket{\downarrow}\otimes \ket{A}\ket{\uparrow}-  \ket{A}\ket{\uparrow}\otimes \ket{A}\ket{\downarrow})\ket{2}+\frac{\sqrt{p(1-p)}}{\sqrt{2}}\Big{[}|A\rangle\ket{\sd}\otimes|B\rangle\ket{\su}-|B\rangle\ket{\su}\otimes|A\rangle\ket{\sd}\nonumber\\
&+&|B\rangle\ket{\sd}\otimes|A\rangle\ket{\su}-|A\rangle\ket{\su}\otimes|B\rangle\ket{\sd}\Big{]}\ket{1} +\frac{p}{\sqrt{2}} (\ket{B}\ket{\downarrow}\otimes \ket{B}\ket{\uparrow}- \ket{B}\ket{\uparrow}\otimes \ket{B}\ket{\downarrow})\ket{0}.
\eea
Equation (\ref{transf}) represents the
state of the tripartite system after the interaction with the measurement
apparatus, and before the result of the measurement is actually read. Clearly, and as a result of the interaction, the detector becomes entangled with the fermionic system. According to the discussion below Eq. (\ref{projected}), projection of $|\Psi\rangle$ onto the state $\ket{1}$ of the detector results in an entangled two-fermion pure state. If no projection is performed, the reduced two-fermion state $\rho_{ff}=\textrm{Tr}_{\mathcal{M}} \ket{\Psi}\bra{\Psi}$ is a mixed state, whose entanglement can be obtained by direct calculation of the (fermionic) concurrence $C(\rho_{ff})$, which quantifies the entanglement between two fermions whose single-particle Hilbert space has dimension 4, in a general (pure or mixed) state $\rho_{ff}$ \cite{ESBL02}. The resulting concurrence is $C(\rho_{ff})=0$, meaning that if no projection is performed (i.e., before the detector clicks), the fermions do not get entangled. Therefore,
even though the interaction of the fermions with the measuring apparatus is essential for creating the fermionic entanglement, 
this entanglement cannot be actually extracted until the detector clicks on 1. The moral of these observations is that
when dealing with identical parties the measurement of the number
of particles present at a given spatial location must not be considered
as ``neutral" or cost-free regarding fermionic entanglement. The fact that entanglement
can be created in these processes is hardly more surprising than the fact
that entanglement can be created when measuring the Bell operator on
a two-qubit system.\\

\section{Entanglement extraction in $N$-fermion systems}
\label{EENfermions}

The process just presented for generating useful entanglement from an initial Slater determinant of two electrons can be generalized to systems composed of $N$ identical fermions in a double-well potential, as can be seen in detail below. An initial Slater determinant of the form \eqref{slater}, corresponding to a situation in which all $N$ fermions are in the well $A$, transforms into another Slater determinant (separable state) once the splitting transformation is performed. However, projection of the evolved state onto states of fixed number of particles in each well (e.g., $M$ in mode $A$ and $N-M$ in mode $B$, with $1\le M \le N-1$), leads to a state that is entangled in the fermionic sense. This fermionic entanglement created in the whole process  (splitting plus detection)  can be evaluated using the measure introduced in \cite{MBVHP2015}. When Alice and Bob operate upon the internal degrees of freedom of the fermions, they share a state that has the same amount of entanglement than the Slater correlations, but in the usual (distinguishable-party) sense,  a result similar to the one obtained for bosons in  \cite{Plenio}.\\

As a first step to investigate this generalization we decompose a Slater determinant into its Schmidt  form. The  Slater determinant can be written as
\bea
\label{AntisymmetricState}
\ket{\psi^{\textit{sl}}} &=& \frac{1}{\sqrt{N!}}\!\!\!\! \sum_{\substack{\{i_1,\ldots,i_N\}\\ \in S_p(1,\ldots,N)}} \!\!\!\!\!\varepsilon^{i_1\ldots i_N} \ket{i_{1}, i_{2}, \ldots, i_{N}}\nonumber\\
&=& \hat f_{1}^\dagger \cdots \hat f_{N}^\dagger \ket{0} = \frac{1}{\sqrt{N!}} \mathcal{A}(\ket{1,2,\ldots,N}),
\eea
where we resorted to the second quantization notation, in which $\hat f^{\dag}_i$ is the fermionic creation operator \cite{ESBL02}, and $\mathcal{A}$ denotes an (unnormalized) antisymmetric vector, defined according to Eq. \eqref{AntisymmetricState}.

In order to write $\ket{\psi^{\textit{sl}}}$ in its Schmidt form, we notice that it can be expressed as

\bea
\label{SlaterProduct}
\ket{\psi^{\textit{sl}}} &=& \frac{1}{\sqrt{N!}}\sum_{\substack{\{i_1,\ldots,i_N\}\\ \in S_p^{(M)}(1,\ldots,N)}} \varepsilon^{i_1\ldots i_N} \mathcal{A}(\ket{i_{1}, i_{2}, \ldots, i_{M}}) \mathcal{A}(\ket{i_{M+1}, \ldots, i_{N}}),
\eea
where $S_p^{(M)}(1,\ldots,N)$ denotes the $\binom{N}{M}$ different ways of choosing the first $M$ indices with $i_1<\cdots<i_M$ from the set $\{1,\ldots,N\}$. The condition $i_{M+1}<...<i_N$ sets the remaining $N-M$ indices of the second antisymmetric product state of Eq.~\eqref{SlaterProduct}. We now write
\bea
\label{As}
\frac{1}{\sqrt{n!}}\mathcal{A}(\ket{i_{1}, i_{2}, \ldots, i_{n}})=\ket{\psi^{\textit{sl}}_{i_1,\ldots,i_n}},
\eea
so that Eq. (\ref{SlaterProduct}) decomposes into its Schmidt form

\bea
\label{SchmidtDecomp}
\ket{\psi^{\textit{sl}}} =\!\!\!\!\!\! \sum_{\substack{\{i_1,\ldots,i_N\}\\ \in S_p^{(M)}(1,\ldots,N)}}\!\!\!\!\!\! \alpha_{i_1\ldots i_N} \ket{\psi^{\textit{sl}}_{i_1,\ldots,i_M}} \ket{\psi^{\textit{sl}}_{i_{M+1},\ldots,i_N}},
\eea
with
\bea
\label{alpha}
\alpha_{i_1\ldots i_N}=\binom{N}{M}^{-\frac{1}{2}}\varepsilon^{i_1\ldots i_N}
\eea
being the Schmidt coefficients, and $\ket{\psi^{\textit{sl}}_{i_1,\ldots,i_M}}\ket{\psi^{\textit{sl}}_{i_{M+1},\ldots,i_N}}$ their corresponding eigenvectors. In this way an $N$-fermion Slater determinant decomposes into the sum of products of $M$- and $(N-M)$- fermion Slater determinants, that result from a particular bipartition $(M:N-M)$ of the complete system. For such a bipartition, the Schmidt rank of a single $N$-fermion Slater determinant is thus $\mathcal{S}^{(M)}_{\textit{sl}}= \binom{N}{M}$. From now on we shall denote by ${\mathcal S}^{(M)}_{\phi}$ the Schmidt rank associated with the $(M: N-M)$ partition effected upon the $N$-fermion system in the state $| \phi \rangle$  (${\mathcal S}^{(M)}_{sl}$ corresponding to the particular case of a global Slater state).

We now consider an initially separable (Slater determinant) state in which all $N$ fermions are in the well $A$. Thus what we previously called state $\ket{i_n}$ will be substituted by $\ket{A_{i_n}}=\ket{A}\ket{i_n}$. The splitting operation associated to the unitary transformation in Eq.~(4) corresponds to the following map involving the fermionic creation operators,
\bea
\hat f^\dagger_{A_{i_n}} \rightarrow  \sqrt{(1-p)} ~\hat f^\dagger_{A_{i_n}} + \sqrt{p} ~\hat f^\dagger_{B_{i_n}}.
\eea
Under this operation the initial state
\bea
\ket{\psi_\text{init}}= \ket{\psi^{\textit{sl}}_{A_1,\ldots,A_N}} = \hat f^\dagger_{A_{1}} \hat f^\dagger_{A_{2}} \ldots \hat f^\dagger_{A_{N}}\ket{0}
\eea
transforms into
\bea
\label{PsiFin}
\ket{\psi_\text{final}} &=& \sum_{M=0}^N \sqrt{(1-p)^M p^{N-M}} 
\!\!\!\!\!\!\sum_{\substack{\{i_1,\ldots,i_N\}\\ \in S_p^{(M)}(1,\ldots,N)}}\!\!\!\!\!\!\!\!\!\!\! \varepsilon^{i_1\ldots i_N} \hat f^\dagger_{A_{i_1}} \ldots \hat f^\dagger_{A_{i_M}} \hat f^\dagger_{B_{i_{M+1}}}\ldots \hat f^\dagger_{B_{i_N}}\ket{0},
\eea
where the sum in the last line runs over the $\binom{N}{M}$ different ways of distributing the $M$ (out of $N$) indices $i_1,\ldots,i_M$ (taken from the set $\{1,\ldots,N\})$ among the $A$'s.
The state (\ref{PsiFin}) is a single Slater determinant, hence is non-entangled in the fermionic sense. This follows from the fact that the splitting transformation does not create fermionic entanglement, as discussed below Eq. (4), and can be easily verified by writing the initial state as
\bea
\label{initialA}
\ket{\psi_\text{init}}&=& \frac{1}{\sqrt{N!}} \sum_{\substack{\{i_1,\ldots,i_{N}\}\\ \in S_p(1,\ldots,N)}}\!\!\!\!\!\! \varepsilon^{i_1\ldots i_N} \ket{{A_{i_1},\ldots,A_{i_N}}},
\eea
so that under the splitting operation,
\bea
\ket{A_{i_n}}=\ket{A}\ket{i_n}\rightarrow \ket{\chi_{i_n}}= (\sqrt{(1-p)} \ket{A}+ \sqrt{p} \ket{B})\ket{i_n},\nonumber
\eea
$\ket{\psi_\text{init}}$ transforms into
\bea
\label{initialAbis}
\ket{\psi_\text{final}}&=& \frac{1}{\sqrt{N!}} \sum_{\substack{\{i_1,\ldots,i_{N}\}\\ \in S_p(1,\ldots,N)}}\!\!\!\!\!\!\varepsilon^{i_1\ldots i_N} \ket{{\chi_{i_1},\ldots,\chi_{i_N}}}.
\eea

Let us now project (\ref{PsiFin}) onto states of fixed number of particles in each spatial mode ($M$ in mode $A$, $N-M$ in mode $B$), thus obtaining
\bea
\label{Proj}
\ket{\psi_\text{proj}}&=&\mathcal{N}\!\!\!\!\!\!\!\!\!\!\!\sum_{\substack{\{i_1,\ldots,i_N\}\\ \in S_p^{(M)}(1,\ldots,N)}}\!\!\!\!\!\!\!\!\!\!\! \varepsilon^{i_1\ldots i_N}  \hat f^\dagger_{A_{i_1}}\ldots \hat f^\dagger_{A_{i_M}} \hat f^\dagger_{B_{i_{M+1}}}\ldots \hat f^\dagger_{B_{i_N}}\ket{0} \nonumber\\
&=&\mathcal{N}\!\!\!\!\!\!\!\!\!\!\!\sum_{\substack{\{i_1,\ldots,i_N\}\\ \in S_p^{(M)}(1,\ldots,N)}} \!\!\!\!\!\!\!\!\!\!\!\varepsilon^{i_1\ldots i_N}  \ket{\psi^{\textit{sl}}_{A_{i_1},\ldots,A_{i_M},B_{i_{M+1}},\ldots,B_{i_{N}}}},
\eea
with $\mathcal{N}$ a normalization factor. Since each term in the sum involves one of the $\binom{N}{M}$ different ways of distributing $M$ indices $i_n$ (with $\{i_n\}=\{1,\ldots,N\})$ among the $A$'s (which fixes the remaining $N-M$ indices appearing in the $B$'s), $\ket{\psi_\text{proj}}$ is a linear combination of Slater determinants, each of which differs from any other in at least one pair of creation operators $\hat f^\dagger_{A_{i_n}} \hat f^\dagger_{B_{i_{m}}}$. Hence, the $\binom{N}{M}$ Slater determinants appearing in Eq. (\ref{Proj}) are all orthogonal, which fixes $\mathcal{N}=\binom{N}{M}^{-1/2}$. Using Eq. (\ref{alpha}), $\ket{\psi_\text{proj}}$ rewrites as
\bea
\label{ProjB}
\ket{\psi_\text{proj}}=\!\!\!\!\!\sum_{\substack{\{i_1,\ldots,i_N\}\\ \in S_p^{(M)}(1,\ldots,N)}} \!\!\!\!\!\!\!\!\alpha_{i_1\ldots i_N}  \ket{\psi^{\textit{sl}}_{A_{i_1},\ldots,A_{i_M},B_{i_{M+1}},\ldots,B_{i_{N}}}}.
\eea

In what follows we will demonstrate that, unlike (\ref{PsiFin}), $\ket{\psi_\text{proj}}$ is entangled in the fermionic sense. This can be verified by performing its Schmidt decomposition. To do that, we first resort to Eq. \eqref{SchmidtDecomp} to expand each of the Slater states appearing in the sum (\ref{ProjB}) into a superposition of two Slater determinants with fixed number of particles, namely $M$ for the first one and $N-M$ for the second, thus obtaining
\bea
\label{ProjB1}
\ket{\psi_\text{proj}} &=&\sum_{\substack{\{i_1,\ldots,i_N\}\\ \in S_p^{(M)}(1,\ldots,N)}} \sum_{\substack{\{j_1,\ldots,j_N\}\\ \in S_p^{(M)}(A_{i_1},\ldots,A_{i_M},B_{i_{M+1}},\ldots,B_{i_N})}}   \alpha_{i_1\ldots i_N}\alpha_{j_1\ldots j_N} \ket{\psi^{sl}_{j_1,\ldots,j_M}}\ket{\psi^{sl}_{j_{M+1},\ldots,j_N}}.
\eea
It is worth mentioning that Eq. (\ref{ProjB1}), describing the sum of Slater determinants each one in its Schmidt form, does not represent in general the Schmidt form of the state $\ket{\psi_\text{proj}}$. In order to obtain such a representation we expand the second sum of Eq. (\ref{ProjB1}) into a sum in which each Slater has a fixed number of particles in the modes $A$ and $B$, that is, 
\bea
\label{ProjB2}
\ket{\psi_\text{proj}} &=& \sum_{\substack{\{i_1,\ldots,i_N\}\\ \in S_p^{(M)}(1,\ldots,N)}} \alpha_{i_1\ldots i_N}  
 \sum_{n=0}^M  \sum_{\substack{\{j_1,\ldots,j_M\}\\ \in S_p^{(n)}(i_1,\ldots,i_M)}}\sum_{\substack{\{j_{M+1},\ldots,j_N\}\\ \in S_p^{(M-n)}(i_{M+1},\ldots,i_N)}} \alpha_{j_1\ldots j_N} \nonumber\\ 
&\times&  \ket{\psi^{sl}_{A_{j_1},\ldots,A_{j_n},B_{j_{M+1}},\ldots,B_{j_{2M-n}}}} 
\ket{\psi^{sl}_{A_{j_{n+1}},\ldots,A_{j_M},B_{j_{2M-n+1}},\ldots,B_{j_N}}}, 
\eea
where $n$ denotes the number of particles located in mode $A$ in the first Slater. The indices $i$'s and $j$'s appearing in the other three sums of Eq.~\eqref{ProjB2} can be reordered as follows:
\bea
\label{ProjB3}
\ket{\psi_\text{proj}} &=& \sum_{n=0}^M  \sum_{\substack{\{i_1,\ldots,i_N\}\\ \in S_p^{(n)}(1,\ldots,N)}}  
 \sum_{\substack{\{j_1,\ldots,j_{N-n}\}\\ \in S_p^{(M-n)}(i_{n+1},\ldots,i_N)}} \alpha_{i_1\ldots i_n,j_1\ldots j_{N-n}}  \sum_{\substack{\{k_{1},\ldots,k_{N-M}\}\\ \in S_p^{(M-n)}(j_{M-n+1},\ldots,j_{N-n})}} \alpha_{i_1\ldots i_n,j_1\ldots j_{M-n}, k_1\ldots k_{N-M}} \nonumber\\ 
&\times&  \ket{\psi^{sl}_{A_{i_1},\ldots,A_{i_n},B_{j_{1}},\ldots,B_{j_{M-n}}}}  \ket{\psi^{sl}_{A_{k_{1}},\ldots,A_{k_{M-n}},B_{k_{M-n+1}},\ldots,B_{k_{N-M}}}}. 
\eea
Because of the symmetry, we consider $1\le M\leq N-M$ hereafter. From this condition it follows that the first Slater state in Eq. (\ref{ProjB3}), with $M$ particles,  is already written in the Schmidt basis, since its corresponding Hilbert space has a dimension which is smaller or equal than the one corresponding to the second Slater determinant. Now, for fixed $n$,  we introduce a global index ${\bf i}_n$ to represent the indices $i$'s and $j$'s appearing in the first two sums of Eq. \eqref{ProjB3}. This global index denotes the $\mathcal{M}=\frac{N!}{n!(M-n)!(N-M)!}$ ways of partitioning the $N$ distinct single-fermion states $\{1,\ldots,N\}$ into the three the sets: $\{{i_1},\ldots,{i_n}$\}, $\{{j_{1}},\ldots,j_{M-n}\}$, and $\{{j_{M-n+1}},\ldots,j_{N-n}\}$, with $n$, $M-n$ and $N-M$ elements, respectively. In terms of the new index we write the first Slater determinant in Eq.~\eqref{ProjB3} as \bea
\ket{\psi_{{{\bf i}_n}, (n,M-n)}} \equiv \ket{\psi^{sl}_{A_{i_1},\ldots,A_{i_n},B_{j_{1}},\ldots,B_{j_{M-n}}}}.
\eea
Since the last sum in \eqref{ProjB3} runs only over the indices $k$ of the second Slater, the Schmidt basis of the $(N-M)$-fermion subsystem is

\beq
\label{SchmidtBase}
\ket{\psi_{{{\bf i}_n}, (M-n,N-2M+n)}} \equiv \binom{N-M}{M-n}^{-\frac{1}{2}} \\
\!\!\!\!\!\!\!\!\!\!\!\!\!\! \sum_{\substack{\{k_{1},\ldots,k_{N-M}\}\\ \in S_p^{(M-n)}(j_{M-n+1},\ldots,j_{N-n})}} \!\!\!\!\!\!\!\!\!\!\!\!\!\!\varepsilon^{i_1\ldots i_n j_1\ldots j_{M-n} k_1\ldots k_{N-M}} 
\ket{\psi^{sl}_{A_{k_{1}},\ldots,A_{k_{M-n}},B_{k_{M-n+1}},\ldots,B_{k_{N-M}}}}. 
\eeq
Finally, the Schmidt decomposition of the projected state reads
\bea
\label{finalStateSchmidtForm}
\ket{\psi_\text{proj}} &=& \sum_{n=0}^M  \sum_{\bf i_n} \lambda_{{\bf i}_n}  
 \ket{\psi_{{{\bf i}_n}, (n,M-n)}} \ket{\psi_{{{\bf i}_n}, (M-n,N-2M+n)}}, 
\eea
with
\bea
\lambda_{{\bf i}_n} = \varepsilon^{i_1\ldots i_n j_1\ldots j_{N-n}} \binom{N-M}{M-n}^{\frac{1}{2}} \binom{N}{M}^{-1}.
\eea

In \cite{MBVHP2015}, a pure $N$-fermion state $\ket{\phi}$ was considered, and it was shown that for any bipartition of the form $M:(N-M)$, the $M$-fermion reduced density matrix satisfies $\text{Tr}\rho_{M}^{2}\leq\binom{N}{M}^{-1}$. It was further demonstrated that the equal sign holds if and only if $\ket{\phi}$ is a Slater determinant. On the other hand, by following the same arguments as exposed in \cite{PMD09}, the $M$-fermion reduced density matrix fulfils  $\rho_M^2=\left(\mathcal{S}_{\phi}^{(M)}\right)^{-1} \rho_M$ if and only if the state $\ket{\phi}$ is a single Slater determinant. Therefore, from these two conditions, it follows that the state $\ket{\phi}$ is entangled if and only if its Schmidt rank $\mathcal{S}_\phi^{(M)}$ is larger than $\mathcal{S}_{sl}^{(M)}$.  The state \eqref{finalStateSchmidtForm} is in the Schmidt decomposition form for the bipartition $M:N-M$, and it has Schmidt rank 
\bea
\mathcal{S}_{\text{proj}}^{(M)} = \sum_{n=0}^M \frac{N!}{n!(M-n)!(N-M)!}.
\eea
Since, 
\beq
\mathcal{S}_{\text{proj}}^{(M)} \binom{N}{M}^{-1}= \sum_{n=0}^M \frac{M!}{n!(M-n)!}=
\sum_{n=0}^M \binom{M}{n}=2^M>1,
\eeq 
it holds that $\mathcal{S}_{\text{proj}}^{(M)} > \binom{N}{M}$,  and therefore we conclude that the state $\ket{\psi_\text{proj}} $ is indeed entangled in the fermionic sense. 

On the other hand, the eigenvalues of the reduced ($M$-fermion) density matrix
\bea
\label{rhoM}
\rho_M &=& \text{Tr}_{N-M}\ket{\psi_\text{proj}}\bra{\psi_\text{proj}} \\
&=&  \sum_{n=0}^M  \sum_{\bf i_n} \lambda_{{\bf i}_n}^2 \ket{\psi_{{{\bf i}_n}, (n,M-n)}} \bra{\psi_{{{\bf i}_n}, (n,M-n)}} \nonumber
\eea
 are given by $\lambda_{{\bf i}_n}^2$, and consequently
\bea
\label{Trace}
\text{Tr}\rho_{M}^{2} &=& \sum_{n=0}^M  \sum_{\bf i_n} \lambda_{{\bf i}_n}^4 = \sum_{n=0}^M  \sum_{\bf i_n} \binom{N-M}{M-n}^2 \binom{N}{M}^{-4} \\
&=& \sum_{n=0}^M \frac{N!}{n!(M-n)!(N-M)!} \binom{N-M}{M-n}^2 \binom{N}{M}^{-4} \nonumber.
\eea
With this expression, the amount of fermionic entanglement generated in the whole process (splitting plus detection) can be determined resorting to the measure introduced in \cite{MBVHP2015}.

We now let independent agents in $A$ and $B$ have access to the particles in their corresponding well, so that Alice and Bob have $M$ and $N-M$ particles, respectively. Notice that though Alice's particles are indistinguishable among themselves, they all are distinguishable from Bob's. With this, the $N$-fermion Hilbert space $\mathcal{H}={\mathcal{H}_f}^{\otimes N}$ splits into $\mathcal{H}_A \otimes \mathcal{H}_B$. Therefore, by projecting (\ref{ProjB}) onto states of the form $\ket{A_{j_1},\ldots,A_{j_M}}\otimes \ket{B_{j_{M+1}},\ldots,B_{j_N}}$ we get
\bea
\label{ProjDistinguishable}
\ket{\psi} \!= \!\!\!\!\!\!\!\!\!\!\!\!\!\!\!\sum_{\substack{\{i_1,\ldots,i_N\}\\ \in S_p^{(M)}(1,\ldots,N)}}\!\!\!\!\!\!\!\!\!\!\! \alpha_{i_1\ldots i_N}\!\!  \ket{i_1,\ldots,i_M}_{\!A}\!\!\otimes\!\!\ \ket{i_{M+1},\ldots,i_N}_{\!B}\!,
\eea
where the subindices $A$ and $B$ have the same meaning as those in Eq.~(8). The state $\ket{\psi}$, with $\text{Tr}\rho_{A}^{2} = \binom{N}{M}^{-1}$, thus corresponds to an entangled state shared by two \textit{distinguishable} entities (Alice and Bob), hence it is legitimately entangled in the usual (distinguishable-party) sense.  The entanglement is manifested in the internal degrees of freedom, whose states are represented by the vectors $\{\ket{i_n}\}$.  This effective entanglement between the particles of both modes is the same as the correlations between $M$ and $N-M$ particles due to the antisymmetry of a single-Slater state of $N$ identical fermions, in consonance  with results obtained in \cite{Plenio} for bosonic systems.

\section{Conclusions}
Our aim in the present work was to  examine the seemingly paradoxical fact that useful entanglement can 
be obtained from a  pure state of $N$ identical fermions exhibiting only Slater correlations
(i.e., correlations due purely to antisymmetrization), even though there 
are deep theoretical reasons for considering such a state as non-entangled. 
To that end we performed a critical analysis of an entanglement generating scheme based on a splitting-plus-detection operation
acting on an initial Slater state. Our analysis highlights the role of the detection process. We argue that accessible 
entanglement can be obtained after the projective measurement only because a state entangled in the 
fermionic sense arises as a result of this operation. In fact, we show that no useful entanglement can be obtained in this way without generating at the same time fermionic entanglement.
Moreover, the quantitative amount of entanglement obtained equals
the amount of fermionic entanglement generated. This implies that the entanglement is created during the measurement process; it is not contained in the initial state. These results are fully consistent with the assertion that the correlations exhibited by states described by one single Slater determinant do not constitute a resource in the standard quantum information sense,
so that these states should not be regarded as entangled.   It would be interesting to extend the present analysis to the
case of initial mixed states and to explore the generation, through processes like the ones considered here,  
of other forms of quantum fermonic correlations,
such as the ones advanced in \cite{MZP2013}, based on a  fermonic generalization of the
measurement induced disturbances approach to  quantum correlations proposed by Luo in \cite{L2008}.  
Any further developments along these or related directions will be very welcome.\\

\begin{acknowledgements}
A.P.M, and P.A.B. acknowledge the Brazilian agencies MEC, MCTI, CAPES, CNPq, and FAPs for the financial support through the \textit{BJT Ci\^encia sem Fronteiras} Program. P.A.B. acknowledges support from the Spanish project grants FIS2014-59311-P (cofinanced by FEDER). A.V.H. gratefully acknowledges financial support from DGAPA, UNAM through project PAPIIT IA101816.
\end{acknowledgements}

\bibliographystyle{apsrev}
\bibliography{Master_Bibtex}

\begin{thebibliography}{}




\bibitem{GMW02} G.C. Ghirardi, L. Marinatto, T. Weber, J. Stat. Phys. \textbf{108}, 49 (2002).

\bibitem{ESBL02} K. Eckert, J. Schliemann, D. Bruss, M. Lewenstein, Ann. Phys. (N.Y.) {\bf 299}, 88 (2002).

\bibitem{PZ02} P. Zanardi, Phys. Rev. A \textbf{65}, 042101 (2002).

\bibitem{YS03} Y. Shi, Phys. Rev. A \textbf{67}, 024301 (2003).

\bibitem{GM04} G.C. Ghirardi, L. Marinatto, Phys. Rev. A \textbf{70}, 012109 (2004).

\bibitem{LNP05} P. Levay, S. Nagy, and J. Pipek, Phys. Rev. A \textbf{72}, 022302 (2005).

\bibitem{GM2005}  G.C.  Ghirardi,  L.  Marinatto, 
Optics and Spectroscopy {\bf 99}, 386 (2005).

\bibitem{NV07} J. Naudts, T. Verhulst, Phys. Rev. A \textbf{75}, 062104 (2007).

\bibitem{PMD09} A. Plastino, D. Manzano, and J. Dehesa, Europhys. Lett. \textbf{86}, 20005 (2009).

\bibitem{BPCP08} A. Borras, A.R. Plastino, M. Casas, A. Plastino, Phys. Rev. A {\bf 78}, 052104 (2008).

\bibitem{ZP10} C. Zander, A.R. Plastino, Phys. Rev. A {\bf 81}, 062128 (2010).

\bibitem{AFOV08} L. Amico, R. Fazio, A. Osterloh, V. Vedral, Rev. Mod. Phys. \textbf{80}, 517 (2008).


\bibitem{PY01} R. Pa\v{s}kauskas and L. You, Phys. Rev. A \textbf{64}, 042310 (2001).

\bibitem{GM05} A.D. Gottlieb and N.J. Mauser, Phys. Rev. Lett. \textbf{95}, 123003 (2005).

\bibitem{DDW06} M.R. Dowling, A.C. Doherty, and H.M. Wiseman, Phys. Rev. A \textbf{73}, 052323(2006).

\bibitem{LZLL01} Y. S. Li, B. Zeng, X. S. Liu, and G. L. Long, Phys. Rev. A \textbf{64}, 054302 (2001).

\bibitem{TMB11} M. C. Tichy, F. Mintert, A. Buchleitner, J. Phys. B: At. Mol. Opt. Phys. {\bf 44}, 192001 (2011).

\bibitem{KO2014} P. Koscik,  A. Okopinska,
 Few-Body Syst.  {\bf 55},  1157 (2014).

\bibitem{ELD2015}  R.O.  Esquivel, S.  Lopez-Rosa, J.S. Dehesa, 
EPL   {\bf  111} ,  40009  (2015). 

\bibitem{MELD2015}    M. Molina-Espiritu,  R.O.  Esquivel,  S. Lopez-Rosa,  J.S.  Dehesa, 
Journ. Chem. Theo. Comp. {\bf  11},   5144 (2015).

\bibitem{SN2015} N.S. Simonovic, R.G.  Nazmitdinov, 
Phys. Rev. A  {\bf  92} , 052332   (2015).
 
\bibitem{Plenio} N. Killoran, M. Cramer, M. B. Plenio, Phys. Rev. Lett {\bf 112}, 150501 (2014).

\bibitem{CMMTF07} D. Cavalcanti, L. M. Malard, F. M. Matinaga, M. O. Terra Cunha, M. F. Santos, Phys. Rev. B \textbf{76}, 113304 (2007).

 \bibitem{SLM01} J. Schliemann, D. Loss, and A. H. MacDonald, Phys. Rev. B {\bf 63}, 085311 (2001).

\bibitem{Strzys2010} M. P. Strzys and J. R. Anglin, Phys. Rev. A {\bf 81}, 043616 (2010).


\bibitem{Tichy2013} M. C. Tichy, J. Phys B: At. Mol. Opt. Phys. {\bf 47}, 103001 (2014).

\bibitem{VHMP2015} A. Vald\'es-Hern\'andez, A. P. Majtey, A. R. Plastino, Phys. Rev. A {\bf 91}, 032313 (2015).

\bibitem{MBVHP2015} A. P. Majtey, P. A. Bouvrie, A. Vald\'es-Hern\'andez, A. R. Plastino,  Phys. Rev. A {\bf 93}, 032335 (2016).

\bibitem{MZP2013}  A.P. Majtey,  C. Zander,  and A.R. Plastino,  
Eur. Phys. Journ. D  {\bf  67},  79  (2013). 

\bibitem{L2008} S. Luo, Phys. Rev. A \textbf{77}, 022301 (2008).

\end{thebibliography}

\end{document}